**Electron bunch generation from a plasma photocathode**

A. Deng[1*], O. Karger[2*], T. Heinemann[2,3,4,5], A. Knetsch[5], P. Scherkl[3,4], G. G. Manahan[3,4], A. Beaton[3,4], D. Ullmann[3,4], G. Wittig[2], A. F. Habib[3,4], Y. Xi[1], M. D. Litos[6], B. D. O'Shea[7], S. Gessner[7], C. I. Clarke[7], S. Z. Green[7], C. A. Lindstrøm[8], E. Adli[8], R. Zgadzaj[9], M. Downer[9], G. Andonian[1,10], A. Murokh[10], D. L. Bruhwiler[11], J. R. Cary[12], M. J. Hogan[7], V. Yakimenko[7], J. B. Rosenzweig[1], B. Hidding[3,4]

[1] Department of Physics and Astronomy, University of California Los Angeles, Los Angeles, California 90095, USA.

[2] Department of Experimental Physics, University of Hamburg, Hamburg 22761, Germany.

[3] Scottish Universities Physics Alliance, Department of Physics, University of Strathclyde, Glasgow G4 0NG, UK.

[4] Cockcroft Institute, Sci-Tech Daresbury, Keckwick Lane, Daresbury, Cheshire WA4 4AD, UK.

[5] Deutsches Elektronen-Synchrotron DESY, Hamburg 22607, Germany.

[6] Center for Integrated Plasma Studies, University of Colorado Boulder, Boulder, Colorado 80303, USA.

[7] SLAC National Accelerator Laboratory, Menlo Park, California 94025, USA.

[8] Department of Physics, University of Oslo, Oslo 0316, Norway.

[9] Department of Physics, University of Texas at Austin, Austin, Texas 78712, USA.

[10] RadiaBeam Technologies, Santa Monica California 90404, USA

[11] RadiaSoft LLC, Boulder, Colorado 80304, USA

[12] Tech-X Corporation, Boulder, Colorado 80303, USA

* These authors contributed equally to this work.

**Plasma waves generated in the wake of intense, relativistic laser[1,2] or particle beams[3,4] can accelerate electron bunches to giga-electronvolt (GeV) energies in centimetre-scale distances. This allows the realization of compact accelerators having emerging applications, ranging from modern light sources such as the free-electron laser (FEL) to energy frontier lepton colliders. In a plasma wakefield accelerator, such multi-gigavolt-per-metre (GV m$^{-1}$) wakefields can accelerate witness electron bunches that are either externally injected[5,6] or captured from the background plasma[7,8]. Here we demonstrate optically triggered injection[9,10,11] and acceleration of electron bunches, generated in a multi-component hydrogen and helium plasma employing a spatially aligned and synchronized laser pulse. This "plasma photocathode" decouples injection from wake excitation by liberating tunnel-ionized helium electrons directly inside the plasma cavity, where these cold electrons are then rapidly boosted to relativistic velocities. The injection regime can be accessed via optical[11] density down-ramp injection[18,19,20], is highly tunable and paves the way to generation of electron beams with unprecedented low transverse emittance, high current and 6D-brightness[12]. This experimental path opens numerous prospects for transformative plasma wakefield accelerator applications based on ultra-high brightness beams.**

The advent of photoinjectors in state-of-the-art linear accelerators (linacs) has enabled the substantial increases in electron beam quality that have ushered in an era of new scientific capabilities, as exemplified by the introduction of the hard X-ray FEL[13]. These photoinjectors produce electron beams in electric fields of ~100 megavolts-per-metre (MV m$^{-1}$). This injection environment largely determines key beam qualities such as the transverse emittance (phase space area) and beam brightness. The strong accelerating field restricts emittance dilution and pulse lengthening by quickly increasing the relativistic Lorentz factor of the beam, $\gamma = (1-v^2/c^2)^{-1/2}$ , where $v$ is the electron velocity and c is the speed of light, thus diminishing these effects that arise from space charge forces[14] and scale as $\gamma^{-2}$. As the electric fields in a plasma wakefield accelerator (PWFA) can exceed those of photoinjectors by more than two orders of magnitude, the PWFA is an attractive environment for high brightness electron beam creation.

A scenario termed plasma photocathode aims to liberate electrons directly inside such plasma wakes via tunneling ionization of neutral gas species at the tight focus of a dedicated, low-power laser pulse. These electrons therefore have low transverse momentum and small initial phase space volume, which is why this injection, which is decoupled from the rapid acceleration process *per se*, permits beams with dramatically improved quality. The process is freely tunable by adjusting injection laser parameters and neutral gas density independent of the acceleration process. It allows transfer of energy from a relatively low-quality driver electron beam into an injected witness bunch of much higher quality. In this way it offers a path to ultrahigh brightness, exceeding the state-of-the-art by many orders of magnitude.

We unlocked the plasma photocathode regime by discovery of a transition process from optical plasma density down-ramp injection[11] in an electron beam-driven plasma wave. This was realized at the Facility for Advanced Accelerator Experimental Tests (FACET) at the SLAC National Accelerator Laboratory[15]. A pre-ionized plasma channel is formed in a hydrogen-helium gas mixture by focusing an ~800 nm Ti:sapphire laser pulse with an axilens[16,17] (see Extended Data Figure 1). The resulting laser intensity distribution with peak intensity of $I_{\mathrm{pre}} \sim 3 \times 10^{14}$ W cm$^{-2}$ selectively tunnel-ionizes hydrogen but not helium, generating a tens-of-centimetre long hydrogen plasma channel of electron density $n_{\mathrm{e,H}} \approx 1.3 \times 10^{17}$ cm$^{-3}$ with limited varying width up to 130 µm.  This channel accommodates a dark-current-free[21], multi-GV m$^{-1}$ wakefield, driven by intense electron drive beams with root-mean-square (r.m.s.) length of $\sigma_z \approx 30$ µm (see Methods). A second, perpendicular-oriented laser pulse is focused to $I_{\mathrm{inj}} \sim 10^{15}$ W cm$^{-2}$ by an off-axis parabolic mirror to release helium electrons from the dual-component gas medium to achieve injection. If this laser pulse crosses the electron driver propagation axis before the electron beam, the plasma wave encounters a plasma density spike due to the presence of the additional localized helium plasma. If this density spike is pronounced enough, it distorts the plasma wave substantially, leading to the injection of hydrogen and helium electrons originating from outside the plasma wave as shown in Figure 1a-c. This is optically-triggered down-ramp injection[18,19,20], which we term plasma torch injection. In contrast, if the laser pulse arrives at exactly the right time – slightly after the driver beam – it releases helium electrons directly within the plasma wave, as shown in Figure 1d-f. This is the plasma photocathode injection mode, which does not require plasma wave perturbation, and in fact profits from lower laser intensities as these produce colder electron populations – direct beam implementation is the only change introduced by the laser in this case.

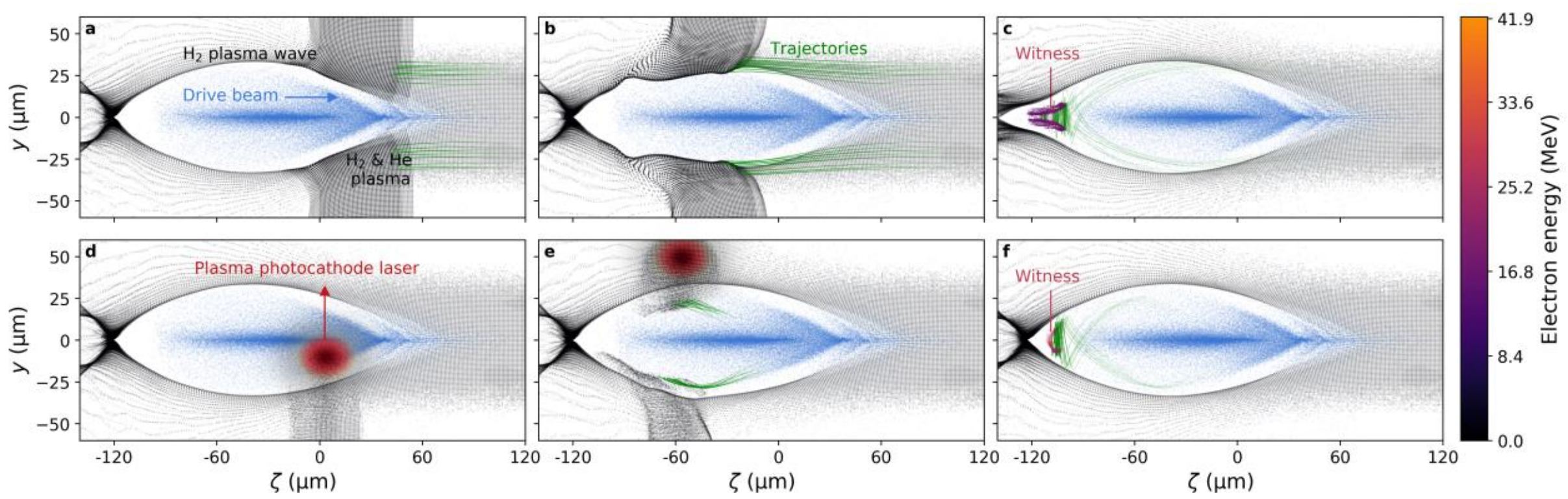

**Figure 1 | Three-dimensional particle-in-cell (PIC) simulations of two injection modes in the beam-driven plasma wave. a-c**, sequential snapshots of the plasma torch injection process, where the driver beam electrons (blue dots) move to the right in the co-moving frame ($\zeta$=$z$-c$t$) and interact with pre-ionized hydrogen plasma channel electrons (colour-coded energy) and a perpendicular plasma filament of ionized hydrogen and helium. This plasma density spike, generated by a 5 mJ laser pulse, distorts the nonlinear plasma "blowout" shape and triggers injection of electrons from outside the blowout, as indicated by selected trajectories (green lines). **d-f,** sequential snapshots of the plasma photocathode injection process. A 0.5 mJ laser pulse (red) releases helium electrons via tunneling ionization inside the hydrogen-based plasma

wave. Selected electron trajectories (green lines) show that the injected electrons originate from inside the plasma blowout. Snapshots **c** and **f** are taken when the respective witness bunches are fully formed, respectively. $\zeta$=0 is defined by the centre of the electron driver beam in the co-moving frame. Only particles within the central slice -2.5 µm < $x$ < 2.5 µm are shown.

Experimentally, an independently controllable fraction of the laser, having pulse duration of 65 fs full-width at half maximum (FWHM) and tunable energy, was focused to a spot size of $w_0 \approx$ 20 µm. Focal intensities of $I_{\text{inj}} > 10^{15}$ W cm$^{-2}$ are achieved with millijoule (mJ) energies and ionize a narrow filament of helium that intersects the driver beam propagation axis at 90° (see Methods). Plasma torch injection is relatively insensitive to electron-beam-to-laser timing, and is achieved for a sufficiently dense plasma filament as long as it spatially overlaps with the driver electron beam and is generated a few femtoseconds up to hundreds of picoseconds or more prior to driver beam arrival. Density down-ramp injection is a promising injection method in its own right but it had not yet been accomplished in beam-driven plasma accelerators before. Here we found it in a straightforward manner due to the steep density gradient[19] obtainable by optically generated down-ramps. Figure 2 demonstrates laser-triggered plasma torch injection by switching the laser on and off repeatedly at 5 mJ over a consecutive series of shots. Experimental evidence of injected electrons is recorded in two ways: as excess electron charge downstream of the plasma source and by the appearance of an energetically distinct grouping of electrons at the spectrometer (see Methods).

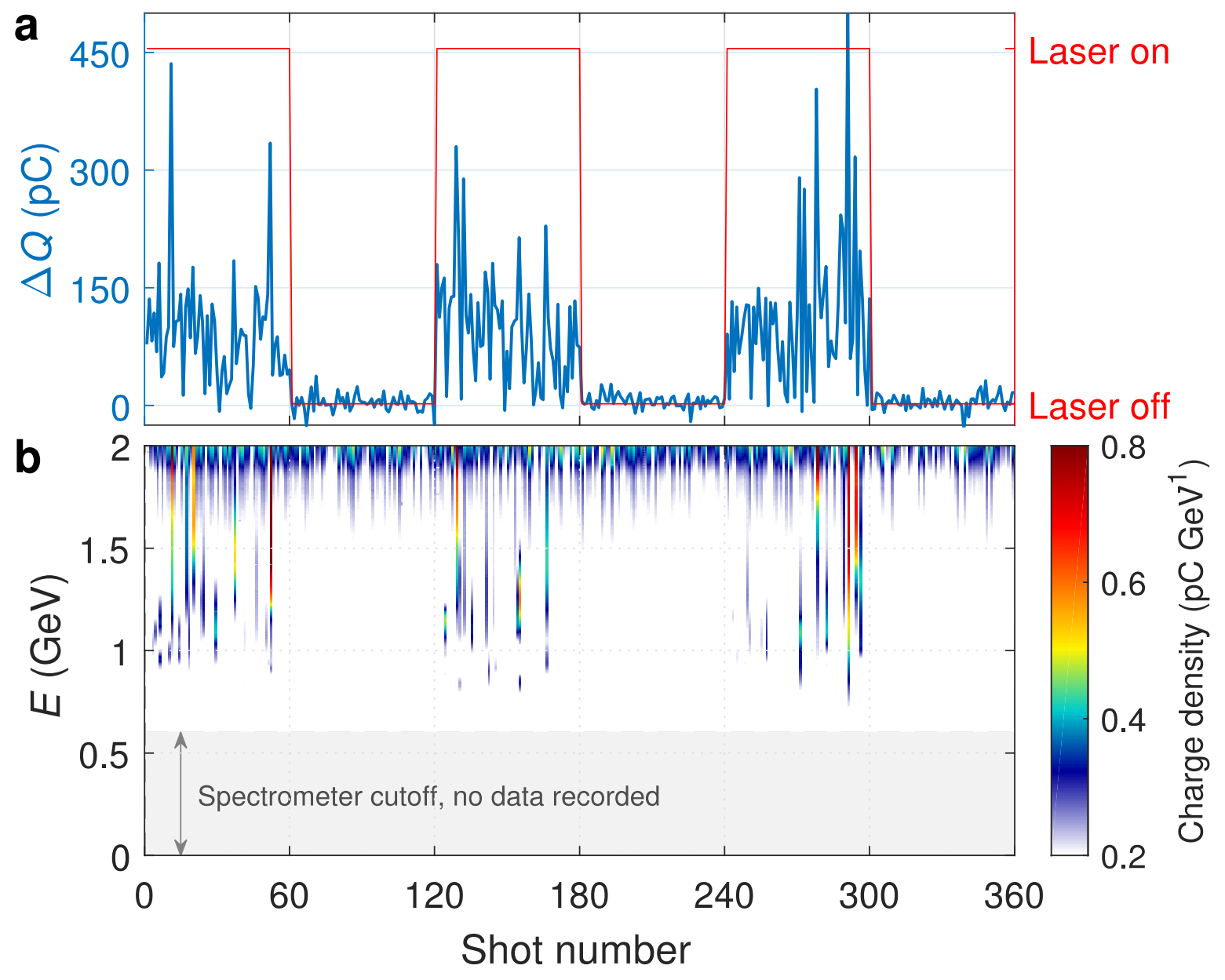

**Figure 2 | Electron charge and spectra obtained from the plasma torch injection process over 360 consecutive shots. a,** injected witness beam charge measured as charge difference at beam position monitors before and after the plasma source. The injection laser pulse arriving 1 ps ahead of the electron driver beam is switched on and off repeatedly every 60 shots, as indicated by the red trace. **b,** corresponding electron spectra measured on the downstream magnetic imaging spectrometer, with optics set to image low energy electrons. The features at the top of the spectra are the decelerated driver beam electrons.

With spatial alignment established via observation of the plasma torch injection, a timing delay scan was performed to determine when the injection laser arrives too late to generate the helium plasma filament prior to arrival of the plasma wave. Beyond this time, plasma torch injection is no longer accessible, as verified by the electron charge and energy diagnostics. This procedure decouples the task of micrometer-precision spatial alignment of the injector laser from femtosecond-level temporal synchronization additionally required for the plasma photocathode. When the laser energy is reduced to a certain level, plasma torch injection also ceases to function because the plasma blowout is not sufficiently deformed by the induced density perturbation. In contrast, helium electrons are trapped even at reduced laser energies when the laser pulse releases them directly inside the plasma blowout, provided a sufficiently high trapping potential exists (see Methods). These transitions are the key to access the plasma photocathode regime and allow to deploy the two injection modes sequentially: starting from plasma torch injection as stepping stone, a timing delay scan, combined with reduced injection laser energy, isolates and reveals the plasma photocathode process as summarized in Figure 3.

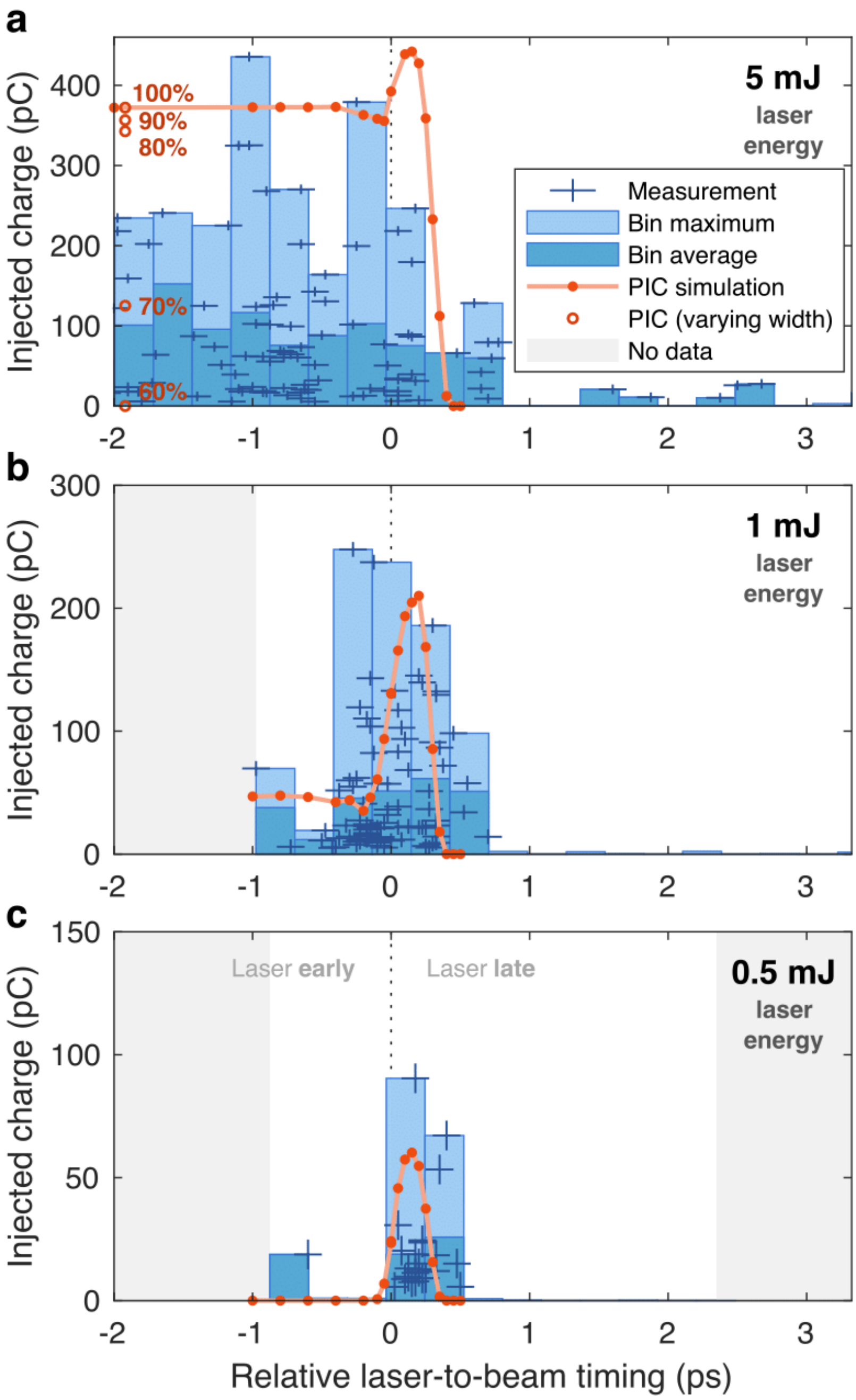

**Figure 3 | Injected charge as a function of laser energy and timing. a,** 5 mJ laser energy (plasma torch injection regime). **b,** 1 mJ laser energy (mixed mode regime). **c**, 0.5 mJ laser energy (plasma photocathode injection regime). The blue crosses represent charge values measured on the downstream spectrometer, and the light blue/dark blue bars indicate the maximum/average charge per time bin, respectively. The corresponding values obtained from PIC-simulations (see Methods) are shown in red. The red open circles in **a** show that the range of injected charge output can be reproduced by variations of channel width and/or alignment. For the grey areas in **b** and **c**, no data was collected. Experimental data are sorted via EOS time stamping to account for the system-inherent time-of-arrival (TOA) jitter (see Methods).

As shown in Fig. 3a, the observed injected charge exhibits a plateau when a 5 mJ laser pulse arrives prior to the driver beam, characteristic of the plasma torch injection mode. When reducing the injection laser energy to 1 mJ, the plasma filament is weakened and down-ramp injection loses its effectiveness. This leads to an intermediate mixed mode regime with reduced injected charge for early laser pulse arrival times and a pronounced peak when the laser pulse arrives immediately after the driver electron beam (Fig. 3b). Finally, a further reduction of laser energy to 0.5 mJ reveals injected charge only in a narrow time window, consistent with the duration of the overlap of the laser with the plasma wave (Fig. 3c). This is the sought-after plasma photocathode injection mode.

An imaging spectrometer downstream of the plasma source allows measurement of electron beam charge, spot size and energy spectrum. Figure 4 presents spectrometer data for representative plasma photocathode witness bunches. Figure 4a shows energy-sorted spectra from ~0.3 GeV to ~0.7 GeV. This range agrees with simulations that investigate details of the acceleration process within the plasma channel (see Methods and Extended Data Figure 2). Close to the spectrometer imaging energy of 0.5 GeV, measurement in the dispersed plane yields a minimum energy spread of 2.1 ± 0.3 % (r.m.s.), as shown in Fig. 4b. Far from this imaging energy, measurement in the non-dispersed plane yields a divergence of 380 ± 30 μrad (r.m.s.), as shown in Fig. 4c. Combining the measured divergence with a calculated beta-function of 1.5 cm at the exit of the plasma results in a minimum normalized emittance of $\varepsilon_n \approx$ 1 mm mrad (see Methods and Extended Data Fig. 3), consistent with simulations.

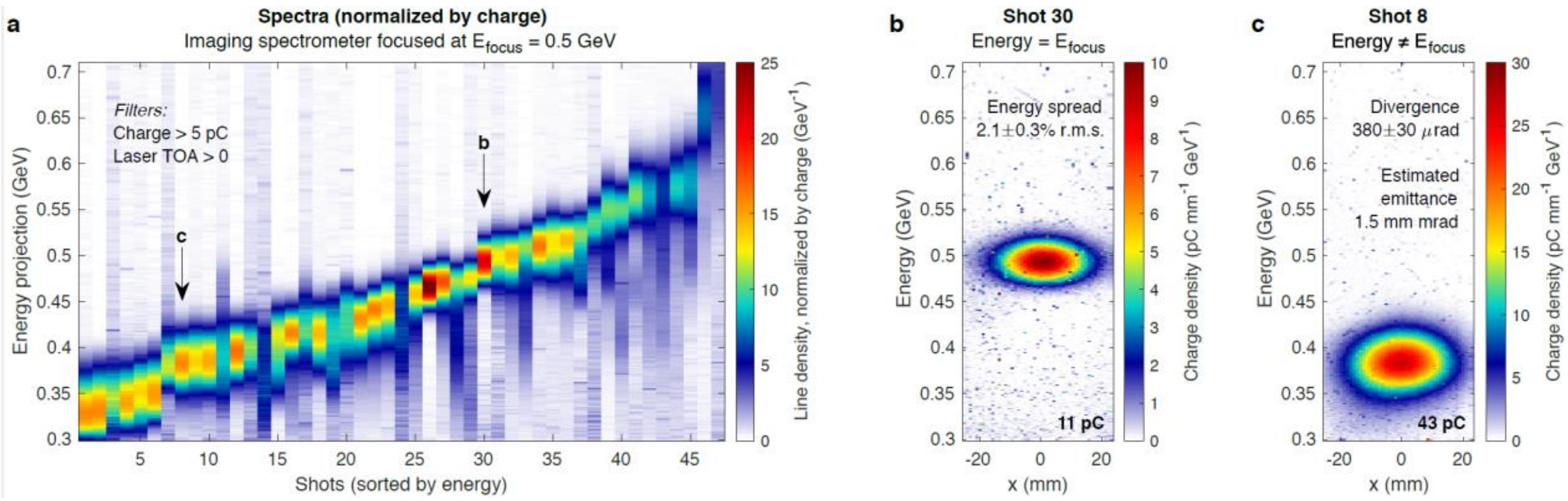

**Figure 4 | Spectrometer data of representative electron bunches measured in the plasma photocathode regime from Fig. 3c. a,** the energy-sorted spectra for shots with charge higher than 5 pC and TOA > 0 with colour-coded charge density. **b,** a selected shot with energy close to imaging energy of 0.5 GeV, showing a minimum energy spread of 2.1 ± 0.3 % (r.m.s.). **c,** a selected shot with energy far from the imaging energy, indicating a horizontal divergence of 380 ± 30 μrad (r.m.s.), and estimated emittance of $\varepsilon_n \approx$ 1.5 mm mrad.

The stability and quality of electron beam production is limited by the plasma channel width, and incoming laser and electron driver beam jitter encountered in the experiment. Simulations show that the plasma channel is a technical bottleneck responsible for injected charge jitter and limited energy gain (see Methods). A wider plasma channel, better pointing stability and an order of magnitude better timing accuracy of driver beam and injection laser are technically feasible and will permit to improve injected charge stability and energy gain substantially. At the same time, a wider plasma channel allows operation at lower plasma densities and in collinear geometry, which also improves beam quality in terms of residual energy spread, emittance and brightness[12].

In summary, we demonstrate controlled electron bunch generation in an electron-driven plasma wakefield accelerator by a decoupled injection laser pulse, releasing electrons at ionization threshold from an overlaid gas component. This allows to realize two complementary injection modes and to seamlessly switch between them: the plasma torch down-ramp injection mode is used to locate and then to access the plasma photocathode regime. The experimental introduction of the plasma photocathode paves the way for production of electron beams with nm-rad-level normalized emittance and brightness up to four orders of magnitude higher than state-of-the-art. Such beams may allow realization of compact, hard x-ray light sources such as free-electron lasers with unprecedented gain[12] characteristics, as well as testing and optimization of emittance preservation for potential future plasma-based linear collider stages.

## METHODS

**Electron driver beam.** The SLAC linac delivers electron beams with a charge up to $Q \approx 3.2$ nC and energy of $W \approx 20.35$ GeV ± 2 % FWHM over a length of approximately 2 km at a repetition rate up to 10 Hz to the FACET experimental area (see Extended Data Fig. 1). The beams are longitudinally compressed in a magnetic chicane to lengths of $\sigma_z \approx$ 20-40 µm (r.m.s.), and are transversally focused by the five quadrupole magnets of the final focus system to $\sigma_x \approx 25$ µm (r.m.s.) and $\sigma_y \approx 30$ µm (r.m.s.), with corresponding beta function of $\beta_x \approx 25$ cm and $\beta_y \approx 100$ cm.

**Plasma source.** The entire ~5.4 m long section between the upstream 50 µm-thick Beryllium foil and a 100 µm-thick diamond foil downstream is filled with a hydrogen/helium gas mixture at densities of $n_{\mathrm{H_2}} \approx n_{\mathrm{He}} \approx 0.65 \times 10^{17}$ cm$^{-3}$. As shown in Extended Data Fig. 1, a Ti:sapphire laser pulse with energy of 170 mJ and pulse duration of 55 fs (FWHM) is focused by an axilens with focal length of 3 m and depth of focus 1 m long, generating a longitudinal Bessel beam intensity profile along the electron driver beam axis. Optical Transition Radiation (OTR) screens are used to align the laser with the electron beam axis. This laser pulse producing an intensity distribution with peak intensity of $3 \times 10^{14}$ W cm$^{-2}$, sufficient to exceed the tunneling ionization threshold of hydrogen, arrives ~20 ps before the electron driver beam. The axially symmetric hydrogen plasma profile produced, as calculated with the Ammosov-Delone-Krainov (ADK) model[22], is shown in Extended Data Fig. 2a. The plasma initiates at the longitudinal position of $z \approx 0.05$ m in Extended Data Fig. 2a, and extends over a length of $\Delta z \approx$ 70 cm. Transversally, it is tapered in a complex manner as result of the Bessel profile, extending to the full width $\Delta r \approx 130$ µm at the widest position. The exponential character of the tunneling ionization rate leads to a sharp transition from gas to fully ionized plasma with peak hydrogen plasma densities of $n_{\mathrm{e,H}} \approx 1.3 \times 10^{17}$ cm$^{-3}$. This choice of density is the result of various considerations. Generally, one would want to work at much lower densities, because then the plasma blowout is much larger, and it is much easier to stably hit the desired position within the blowout with the injection laser for a technically given absolute laser-electron-beam jitter in space and time. This strongly enhances reproducibility of the output beam. At the same time, a transverse kick by the drive beam (see Fig. 1d-f), which increases the transverse emittance of the witness bunch, would be avoided, and a reduced residual energy spread would be obtained[12]. However, the plasma blowout must fit within the preionized plasma channel of limited and varying width. This limit is not fundamental, but was experimentally imposed by the available laser system energy and available space. Even with an experimentally optimized density of $n_{e,H}$ $\approx 1.3 \times 10^{17}$ cm$^{-3}$, the blowout touches the boundaries of the channel for most of the acceleration, which compromises the blowout strength and strongly limits the energy gain (Extended Data Fig. 2a). An upper hydrogen plasma density limit for the given FACET electron beam results from unwanted helium ionization and dark current by the more strongly pinching driver beam fields and wakefields[7,8,21], which sets in approximately at $n_{e,H} > 2 \times 10^{17}$ cm$^{-3}$. It should be noted that in contrast to purely laser-based ionization injection schemes[23,28], the injection in electron beam-driven approaches arises from the independently tunable helium density in the mixture. Adjusting the helium density is therefore an independent knob which can be used to tune the injected charge from femto-Coulomb to nano-Coulomb levels. Laser-based injection schemes such as[23,28-30] are showing transformative impact on optimizing laser-plasma wakefield acceleration, which in turn could lead to production of electron beams employable as drivers for beam-driven plasma wakefield acceleration[9] in ultracompact hybrid systems.

**Probe and plasma photocathode injection lasers.** A laser pulse with duration of 65 fs (FWHM) is split up into two pulses that traverse optical paths which are independently tunable in energy and equipped with delay stages for temporal synchronization. One collimated laser arm serves an upstream electro-optic sampling (EOS) diagnostic for shot-to-shot time-of-arrival (TOA) measurement of driver electron beam-to-laser relative timing, while the plasma photocathode injection laser arm is focused by an *f*/22.9 off-axis-parabolic mirror to a spot size of ~ 20 µm perpendicular to the electron beam axis. The upstream OTR screen is used for spatial alignment of the focused injection laser and the driver beam (see Extended Data Fig. 1). The EOS system non-destructively measures shot-to-shot-jitter between the laser and electron driver beam as 109 ± 12 fs (r.m.s.), and simultaneously records the TOA for the injection experiments with a resolution of 25 fs. A CCD camera beneath the beam line images the laser-generated plasma filament across the electron driver beam axis. These diagnostics combined facilitate the alignment and synchronization of the injection laser and electron-driven plasma wake.

**Electron witness beam generation and measurement.** In an unperturbed plasma wave, the trapping condition[7,23] is given by $(\Psi_{max} - \Psi)/[(m_0c^2/e)(1-\gamma_{ph}^{-1})] < -1$, where $\Psi$ is the electrostatic trapping potential, $m_0$ and e are the electron rest mass and charge, respectively, c is the speed of light and $\gamma_{ph}$ is the relativistic plasma wave Lorentz factor. This condition is easily fulfilled by the strong plasma wave at FACET (see Supplementary Movies). As shown in Extended Data Fig. 2, electrons injected at around $z_{inj} \approx 0.2$ m are rapidly accelerated by tens of GV $m^{-1}$ fields in this comparably wide part of the plasma. Due to the transversely tapered plasma profile, however, the wakefields are progressively reduced in $z$ and the wave geometry elongates. Witness electrons are then partially located in the decelerating phase of the plasma wave, which limits the total energy gain. Extended Data Fig. 2b shows the expected range of final witness electron energies of up to ~1.9 GeV for the given plasma profile with consideration of possible injection positions in the lab frame and the trapping positions in the co-moving frame. It is in good agreement with experimental data, which have been obtained by the imaging spectrometer after transporting the electrons downstream in a well-characterized beam line. The integrated probability of injecting electrons, a measure of expected witness charge, is highest for injection that results in production 0.3-0.7 GeV electrons. A wider pre-ionized plasma channel would be suitable to resolve issues of injected electron jitter and energy gain limitations.

After exiting the plasma and entering the vacuum section through the diamond window (see Extended Data Fig. 1), the electrons are captured and focused by a quadrupole doublet, and dispersed by a dipole magnet onto a CCD-monitored phosphor screen located 22 m after the end of the plasma. Charge is measured by BPMs (with an accuracy of 4-5%) and also by the signal intensity on the phosphor screen (calibrated based on BPM measurements, in combination having an accuracy of ~10%). Extended Data Fig. 3 summarizes the analysis of energy spread and divergence measurements. The energy spread measured on the spectrometer has an additional contribution from the vertical emittance, but close to the imaging energy of 0.5 GeV the emittance contribution is negligible, thus resolving the actual energy spread (see Extended Data Fig. 3a). To derive the divergence, Gaussian fits of the spectral charge density in the transverse (non-dispersive) direction are combined with beam transport simulations including multiple Coulomb scattering[24] in the exit foil. The corresponding emittance is calculated with a beta function of 1.5 cm, obtained by integrating Hill's equation through the modeled plasma density profile. The measured transverse beam sizes and estimated emittances are shown in Extended Data Fig. 3b.

**Plasma wakefield acceleration simulations.** The fully explicit 3D particle-in-cell (PIC) code VSim[25] is used to simulate the driver beam-plasma interaction and the injection processes depicted in Fig. 1, Fig. 3 and the Supplementary Movies. An accurate representation of the pre-ionized plasma profile (see Extended Data Figure 2a) is directly loaded into the simulations. The electron driver beam has been implemented as a tri-Gaussian beam based on the experimental measurements of $\sigma_x$, $\sigma_y$, $\sigma_z$, charge, energy, energy spread and emittance. The beam was initialized at the entrance of the plasma. The simulation box has a size of 500 µm × 300 µm × 300 µm in the longitudinal dimension and the two transverse dimensions, respectively. An additional 16 µm on each transverse border are reserved as absorbing boundaries via perfectly matched layers. The number of the cells for the simulation box is 250 × 166 × 166. The coordinate system represents a co-moving frame at the speed of light. Eight macro-particles per cell (PPC) are used to model the plasma, and 16 PPC for the driver beam, respectively. Time-resolved PIC simulations are needed to resolve the injection processes. Hence, the number of particles and the spatial resolution of the simulations are chosen as a trade-off between desired accuracy and the long propagation distance. All simulations use the Yee scheme for electro-magnetic field updates, where the time-step is chosen to be 0.5 times the Courant limit. Charge and current deposition is applied according to the Esirkepov algorithm in 3rd order, and the Boris method is used to push the macroparticles.

The driver beam and wakefield evolution are simulated throughout the plasma, allowing the mapping of the accelerating field and trapping conditions over the whole propagation distance to estimate the expected energy gain (see Extended Data Figure 2). The configuration at the experimental injection point $z_{\mathrm{inj}} \approx 0.2$ m inside the plasma (see Extended Data Fig. 2a) serves as the initial state of the individual injection simulations shown in Fig. 1 and the Supplementary Movies, as well as for the data points in Fig. 3. Simulations are performed with different laser energies and relative laser-to-beam timing, and the trapped charge in the wake is presented in Fig. 3, where zero is defined as the time when the electron and laser beam centres cross each other.

Three complementary simulation approaches are used to cover the large delay range. For positive timing values, the injection laser is added as an envelope in paraxial approximation, and tunneling ionization is calculated based on an averaged ADK model[26,27]. For negative timing values, the electron driver beam is externally loaded into a longer simulation box, allowing the laser to dynamically generate the plasma filament before the driver beam arrival. For even larger delays (< -1 ps), the fully formed plasma filament is loaded in the simulation. The injection simulations were performed 5 mm beyond the injection point and the same spatial and energy cuts were applied for all cases to discard the low energy and high divergence electrons which would not survive the downstream acceleration and beam transport. All simulation approaches are consistent and in good agreement with the experimentally obtained charge values, as shown in Fig. 3. We note that for the highest energy considered (5 mJ), a perfectly Gaussian laser pulse may even ionize the second level of helium. Simulations including this additional ionization level confirm that the general injection behavior is maintained albeit at increased charge values. This experimental possibility of second level ionization can account for occasionally higher charge values as shown in Fig. 3(a). Supplementary Movies visualize the two injection modes shown in Fig. 1.

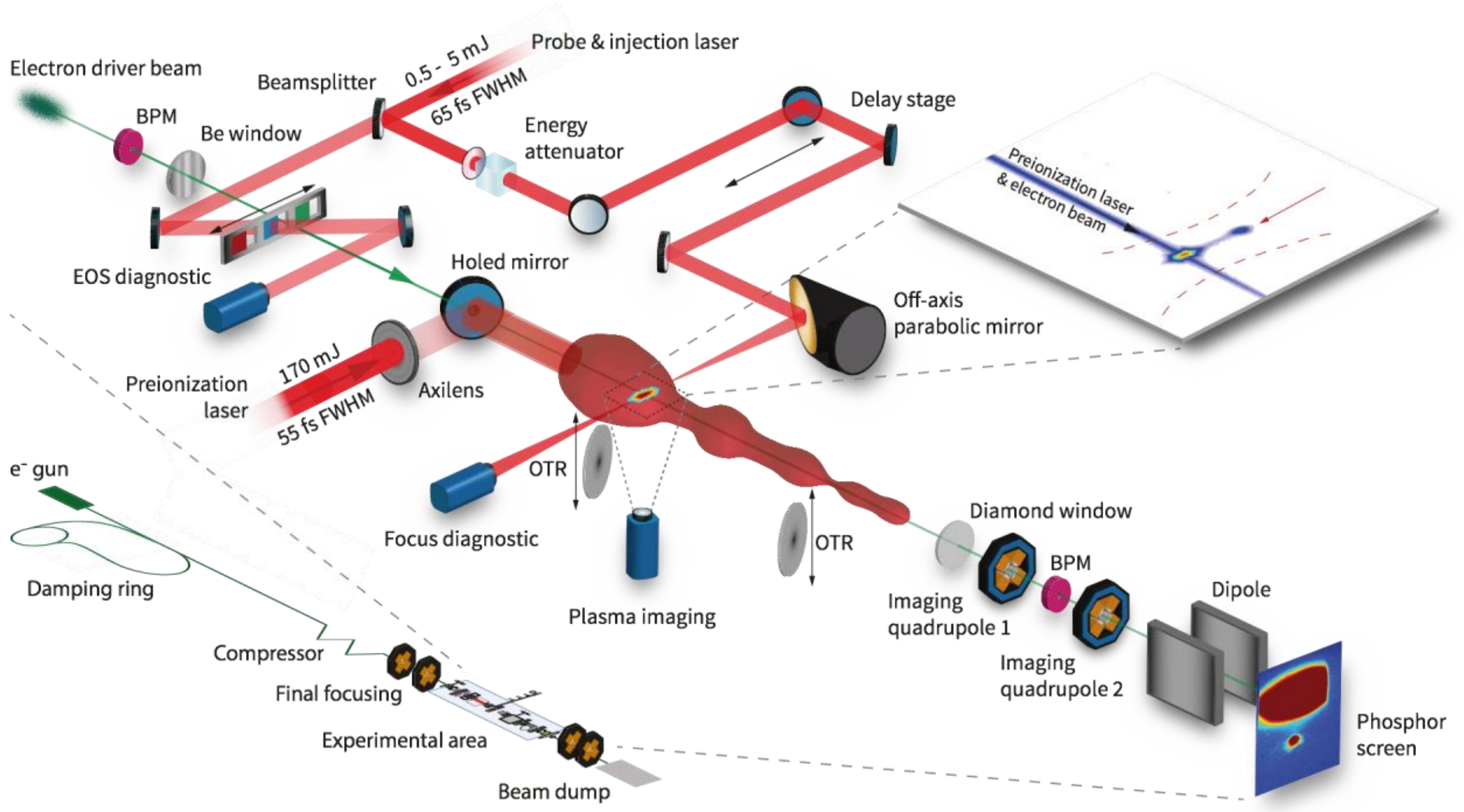

**Extended Data Figure 1 | Overview of experimental setup at SLAC FACET.** The 2 km linac-generated electron driver beam arrives from the top left direction, and a plasma channel is generated by focusing a high-power laser pulse along the electron beam axis with an axilens. A separate laser pulse feeds an electro-optical sampling (EOS) unit upstream of the interaction point, and the plasma photocathode laser is focused onto the electron beam axis at an angle of 90° by an off-axis parabolic mirror. Beryllium (Be) and diamond windows confine the gas to the plasma source region. Optical transition radiation (OTR) screens are used to align the lasers with the electron beam axis. Beam position monitors (BPMs) upstream and downstream of the plasma source are used for charge measurement, and a beam transport line with an imaging spectrometer is used to measure and derive the electron bunch energy, energy spread and divergence.

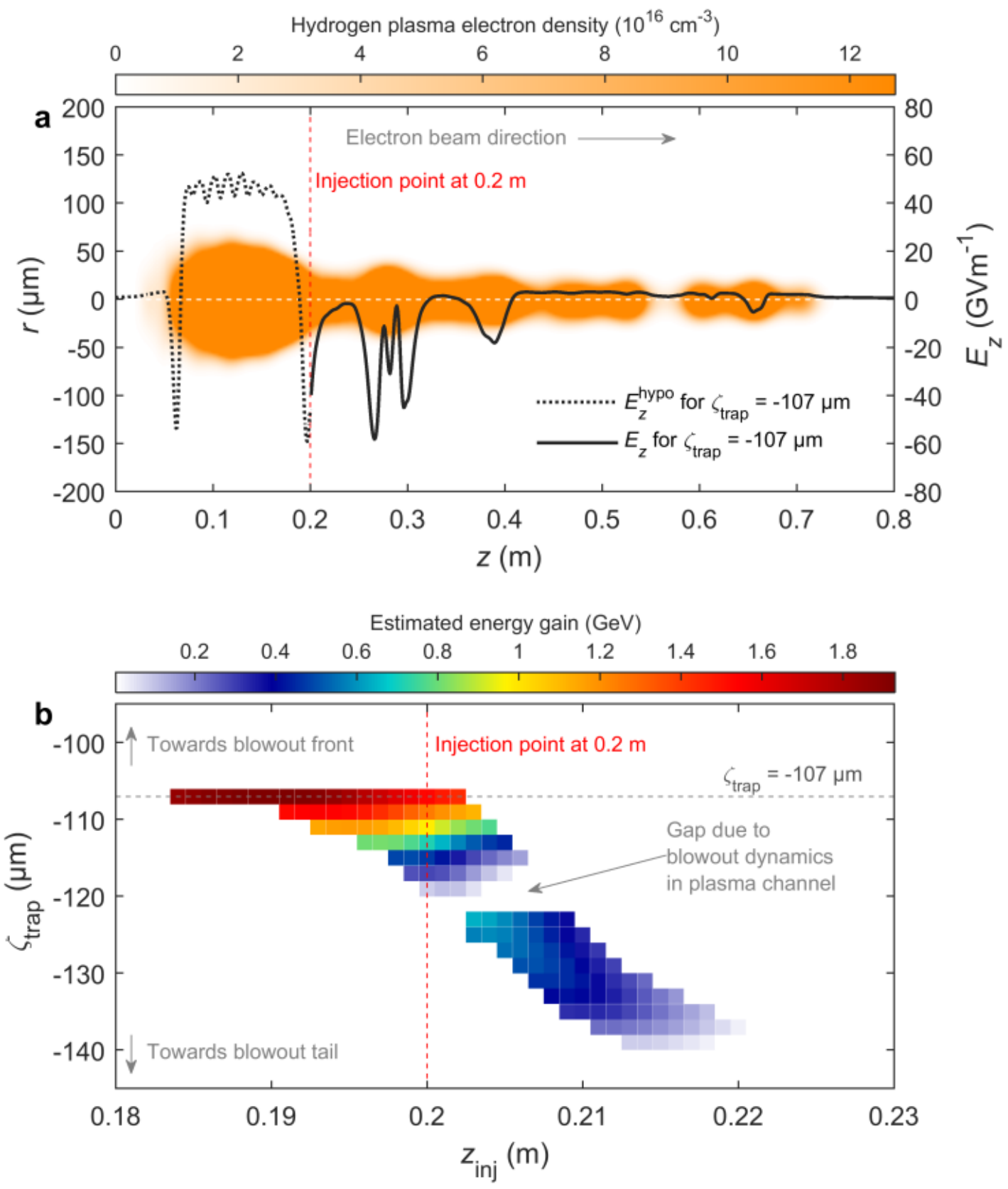

**Extended Data Figure 2 | Calculated plasma source profile, longitudinal electric field and energy range of accelerated electrons. a,** the hydrogen plasma density profile produced by the axilens-focused laser pulse along the electron driver beam axis *z*. The dashed red line at $z \approx 0.2$ m indicates the position of the transverse injection laser pulse in the laboratory frame. The varying width of the plasma source distorts the plasma oscillation trajectories and thus modulates the blowout strength and shape along *z*. This feature leads to a varying longitudinal electric field $E_z$ during the propagation through the plasma; the evolution of the longitudinal field depends on the specific positions $\zeta = z - ct$ of the witness electrons in the co-moving frame. The black line shows the simulated $E_z$ (right *y*-axis) sampled by the electrons trapped at $\zeta_{trap} \approx$ -107 μm during acceleration, where $\zeta = 0$ is defined by the centre of the electron driver beam in the co-moving frame as in Figure 1. **b,** the expected final energy range of the accelerated electrons (colour-coded) arising from different injection positions in the laboratory frame around $z_{inj} \approx 0.2$ m and corresponding possible trapping positions $\zeta_{trap}$ from -107 μm to -

139 µm. In the underlying simulations of the propagating blowout evolution, electrons that move into the defocusing phase of the blowout at any time do not survive the acceleration process, resulting in the white gap. A wider plasma channel allows to harness constantly accelerating electric fields, which then implies tens of GeV energy gain over the length of the plasma channel[6].

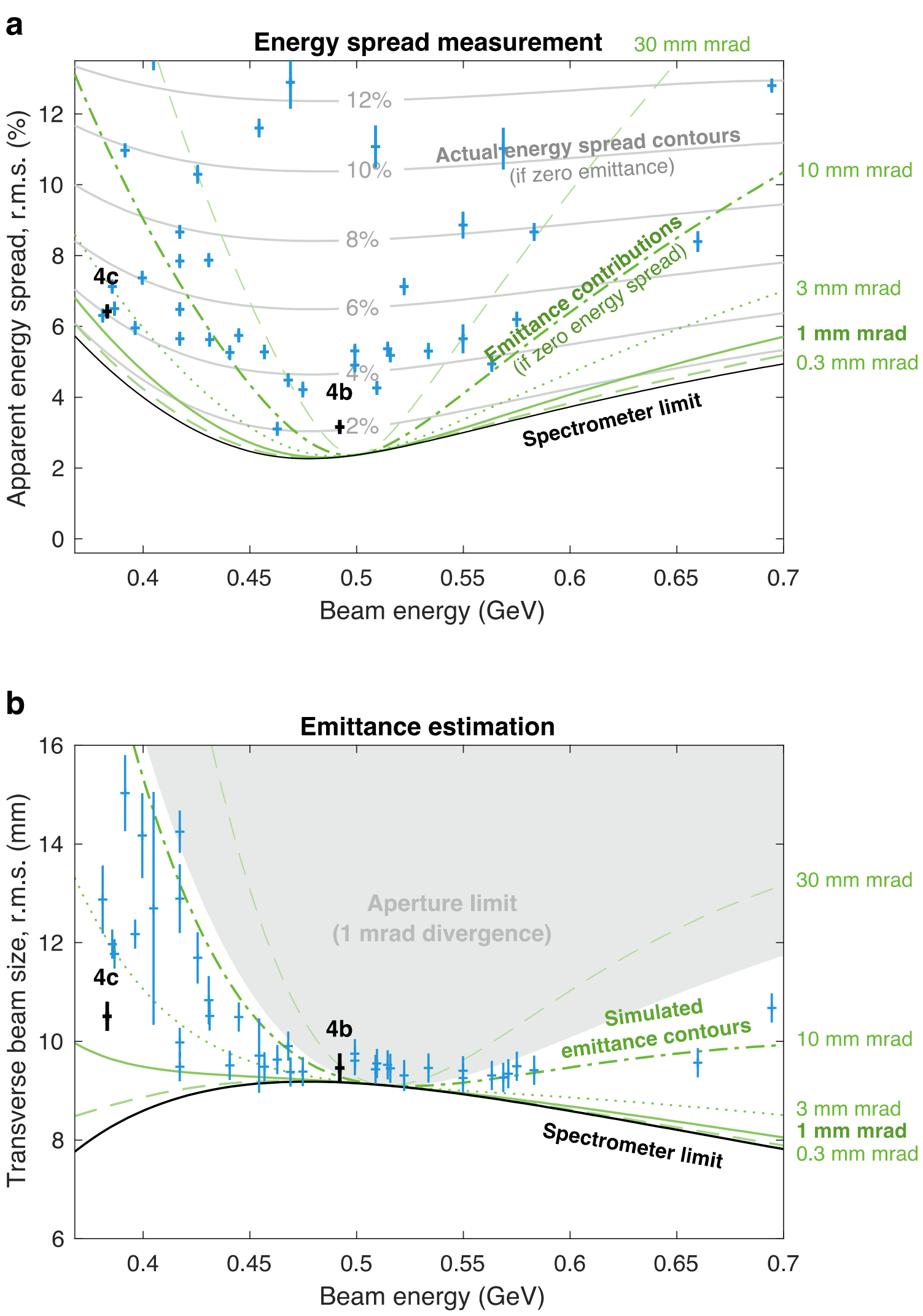

**Extended Data Figure 3 | Statistical analysis of beam quality for the shots in Figure 4 (a).** **a**, the experimentally measured apparent r.m.s. energy spreads (blue data points), as a result of the actual energy spread (grey lines) and an additional vertical emittance contribution (green lines). A minimum energy spread of 2 % is obtained for the shot shown in Fig. 4b. **b**, the

corresponding r.m.s. transverse beam sizes in the non-dispersed plane with emittance estimation (green contours) using a beta function of 1.5 cm. A shot with emittance of 1.5 mm mrad is shown in Fig. 4c. The grey area shows a divergence limit of 1 mrad imposed by the aperture of a laser out-coupling mirror having a central hole for beam passage (located between the second OTR and the diamond window, not shown in Extended Fig. 1). The black lines represent the spectromter limits according to a simulated zero emittance and zero energy spread of the beam. Shots with compromised data characteristics are omitted.

**Data availability**

Data associated with research published in this paper will be available at publication.

**Supplementary Information** is available in the online version of the paper.

## Acknowledgements

The FACET E210 plasma wakefield acceleration experiment was built and operated with support from UCLA (US DOE contract DE-SC0009914), RadiaBeam Technologies (DOE contract DE-SC0009533), the FACET E200 team and the U.S. Department of Energy under contract number DE-AC02-76SF00515, H2020 EuPRAXIA (Grant No. 653782), Helmholtz VI, EPSRC (Grant No. EP/N028694/1), and the Research Council of Norway (Grant No. 230450).

This work used computational resources of the National Energy Research Scientific Computing Center, which is supported by DOE DE-AC02-05CH11231, of JURECA (Project hhh36), of HLRN and of Shaheen (Project k1191). D.L.B. acknowledges support of the US DOE Office of High Energy Physics under Award No. DE-SC0013855.

## Author Contributions

All authors contributed extensively to the work presented in this paper.

## Author Information

Reprints and permissions information is available at www.nature.com/reprints. The authors declare no competing financial interests. Readers are welcome to comment on the online version of the paper. Correspondence should be addressed to bernhard.hidding@strath.ac.uk